\documentclass{aa} 
\usepackage{hyperref}
\usepackage{lscape}

\usepackage{csvsimple}

\usepackage{graphicx}
\usepackage{txfonts}
\begin{document}

   \title{Disentangling the two sub-populations of early Herbig Be stars using VLT/X-Shooter spectra}

   \subtitle{}
   
   \author{B. Shridharan$^{1}$\thanks{E-mail: shridharan.b@res.christuniversity.in},
Blesson Mathew$^{1}$,
R. Arun$^{2}$,
T.B. Cysil$^{1}$,
A. Subramaniam$^{2}$,
P. Manoj$^{3}$,
G. Maheswar$^{2}$,
T.P. Sudheesh$^{1}$}
    \authorrunning{Shridharan et al. }
   \institute{Department of Physics and Electronics, CHRIST (Deemed to be University), Bangalore 560029, India
   \and Indian Institute of Astrophysics, Sarjapur Road, Koramangala, Bangalore 560034, India
   \and Tata Institute of Fundamental Research, Homi Bhabha Road, Mumbai 400005, India}

   \date{Received xxxx; accepted xxxx}

 
  \abstract
   {Early Herbig Be (HBe) stars are massive, young stars accreting through the Boundary Layer mechanism. However, given the rapid ($<$ 2 Myr) evolution of early Herbig stars to the main-sequence phase, studying the evolution of the circumstellar medium around these stars can be a cumbersome exercise.}
   {In this work, we study the sample of early (B0-B5) HBe stars using the correlation between H$\alpha$ emission strength and near--infrared excess, complemented by the analysis of various emission features in the X-Shooter spectra. }
   {We segregate the sample of 37 early HBe stars based on the median values of H$\alpha$ equivalent width (EW) and near--infrared index (n(J$-$H)) distributions. The stars with |H$\alpha$ EW| $>$ 50 \AA~and n(J$-$H) $>$ -2 are classified as intense HBe stars and stars with |H$\alpha$ EW| $<$ 50 \AA~and n(J$-$H) $<$ -2 as weak HBe stars. Using the VLT/X--Shooter spectra of five intense and eight weak HBe stars, we visually checked for the differences in intensity and profiles of various H{\sc I} and metallic emission lines commonly observed in Herbig stars. }
   {We propose that the intense HBe stars possess an inner disk close to the star (as apparent from the high near-infrared excess) and an active circumstellar environment (as seen from high H$\alpha$ EW value and presence of emission lines belonging to Fe{\sc II}, Ca{\sc II}, O{\sc I} and [O{\sc I}]). However, for weak HBe stars, the inner disk has cleared, and the circumstellar environment appears more evolved than for intense HBe stars. Furthermore, we compiled a sample of $\sim$58,000 emission-line stars published in \textit{Gaia DR3} to identify more intense HBe candidates. Further spectroscopic studies of these candidates will help us to understand the evolution of the inner ($\sim$a few au) disk in early HBe stars.  }
  {}

   \keywords{ Stars: emission-line, Be -- circumstellar matter -- methods: data analysis -- techniques: spectroscopic
               }

   \maketitle
%

\section{Introduction}

Herbig Ae/Be stars are pre-main sequence (PMS) stars belonging to A and B spectral types (HAeBe) with their mass ranging from 2 to 10 M$_{\odot}$\citep{herbig1960spectra, strom1972optical} . The optical and infrared (IR) spectra of HAeBe stars are known to show emission lines of various allowed and forbidden transitions such as H$\alpha$, H$\beta$, Ca{\sc II}, O{\sc I}, and [O{\sc I}] \citep{2015FairlambMNRAS.453..976F, hamann1992emission}. In addition, they show excess flux in the infrared wavelengths due to the energy reprocessing by dust grains in the outer disk \citep{hillenbrand1992herbig, cohen1980infrared, malfait1998ultraviolet}. The circumstellar medium of HAeBe stars is active with different emission lines arising from circumstellar disk, accretion columns and disk winds around them (refer \citealp{brittain2023herbig} for a recent review on HAeBe stars). 

However, in recent years, studies have noted the distinction between various properties of HBe and HAe stars. Since the mass varies drastically between B and A spectral types, the processes occurring in the stellar interiors also differ between HBe and HAe stars. The properties of accretion \citep{wichittanakom2020accretion} and variability \citep{villebrun2019magnetic,mendigutia2011optical} have been reported to differ between HAe and HBe stars. One of the significant differences between them is the mode of accretion from the circumstellar disk. HAe and late HBe stars are known to accrete through the Magnetospheric accretion (MA) paradigm (see \citet{mendigutia2020mass} for a review), where the star's magnetosphere truncates the circumstellar disk at a few stellar radii and funnels the material onto the star through the magnetospheric field lines. Due to the absence of a strong magnetic field \citep{gregory2012can}, early HBe stars do not accrete through MA and instead does through Boundary Layer (BL) mechanism. In BL mode, the keplerian circumstellar disk almost reaches the stellar surface and material gets transferred onto the star through the interface between the disk and the stellar surface (known as the boundary layer) \citep{lynden1974evolution}.  

An interesting HBe star to note in the context of BL accretion is MWC 297, which is a nearby ($\sim 400$ pc; \citealp{bailer2021estimating}), massive young star of B1.5V spectral type. This HBe star has been extensively studied and is known to be actively accreting from its circumstellar disk. It is known to have strong Balmer emission (|H$\alpha$ EW| > 200 \AA), double-peaked [OI] emission and double-peaked CO emission, which is suggestive of the presence of an nearly edge-on circumstellar disk \citep{banzatti2022scanning, drew1997mwc, zickgraf2003kinematical, acke2008mwc}. Further, interferometric observations and modelling of MWC 297 did not reveal a cavity or inner gap in the disk \citep{kluska2020family, acke2008mwc}. Thus, MWC 297 accretes material through the BL mechanism. Similar direct pieces of evidence are not available for other early HBe stars. Hence, this demands the study of a large sample of HBe stars, belonging to diverse environments, to evaluate the mode of accretion and to understand the nature of the circumstellar medium.

The motivation for this work was from \citet{shridharan2023hi}, where they performed H{\sc I} line analysis of HAeBe stars using the X-shooter spectra. In addition, they studied the statistical presence of higher-order emission lines of H{\sc I} in HAeBe stars with respect to the stellar parameters and from which, they noticed bimodality in higher-order emission lines in the sample of early (B0-B5) HBe stars. Given that the ages (pre-main sequence lifetime) of early HBe stars are less than 2 Myr and the fact that they already lost most of their inner circumstellar disk, the analysis of these objects is vital to bridge the stages in the intermediate-mass star formation. Using this as a lead, we checked for the differences between the properties of the two populations, identified due to the distinction in the presence of higher-order Balmer emission lines. In the present work, we did further analysis and found the presence of two populations of early HBe stars (B0-B5) in the H$\alpha$ versus near-infrared (NIR) excess space. Since H$\alpha$ and NIR excess in HAeBe are known to be associated with the inner ($\sim$a few au) circumstellar medium \citep{manoj2006evolution}, we suspect a difference in the circumstellar medium of the two populations.  Hence, a study of these two groups can help in understanding the evolution of the inner circumstellar disk in stars undergoing BL accretion. From a search in the literature, we found three instances where two populations of HBe stars are reported. \citet{hillenbrand1992herbig} found that most of the Group III stars (with low NIR excess) belong to B spectral type and \citet{corcoran1998wind} proposed a class of "weak-line" HAeBes for stars with no forbidden line emission and H$\alpha$ equivalent width (EW) < 15\AA. \citet{banzatti2018observing} observed a dichotomy in the NIR excess of their group I sources. The group I sources either showed high (> 25\%) or low (< 10\%) NIR excess. They also found that the strength of NIR excess is correlated with the radius at which CO is detected. The sources with low NIR excess show CO emission at larger radii compared to sources with high NIR excess, which implies that the inner region is depleted in low NIR excess sources. In this work, we report the detection of two populations of early HBe stars and discuss the distinction between their circumstellar medium.
 
The paper is structured as follows. The dataset used in this work is explained in Section 2. The differences in the NIR excess, H$\alpha$ and spectral features between the two populations are described in Section 3. We also evaluate a few interesting questions regarding the identification of two sub-populations of early HBe stars in Section 4, followed by a summary of this work in Section 5.
 
\section{Data used for this study}

We combined the well-studied HAeBe catalogues of \citet{vioque2018gaia} and \citet{guzmandiaz2021} to compile a comprehensive set of stellar parameters. From this combined catalogue, we selected 37 Herbig Be stars with spectral type earlier than B5 (as estimated by \citealp{guzmandiaz2021}) for our analysis.
The NIR magnitudes (J, H and K$_S$) used in the study are the photometric data from Two Micron All Sky Survey (2MASS, \citealp{2mass2006AJ....131.1163S}) and the mid-infrared (MIR) magnitudes (W1, W2 and W3) are taken from Wide-field Infrared Survey Explorer (WISE, \citealp{wisecutri2014yCat.2328....0C}). Further, the medium-resolution spectra of HBe stars observed with the X-Shooter instrument are retrieved from the archive and are used for the present study. X-Shooter is an Echelle spectrograph mounted at the UT3 Cassegrain focus of VLT telescope (Cerro Paranal, Chile), that provides spectra covering a large wavelength range of 3000--23000 \AA, split into three arms and taken simultaneously. The arms are split into the following: the UVB arm from 3000--5600 \AA, the VIS arm from 5500--10200 \AA, and the NIR arm from 10200--24800 \AA. The smallest slits available with widths of 0.5”, 0.4” and 0.4” were used to provide the highest possible resolutions of R $=$ 9700, 18400 and 11600 for the respective UVB, VIS and NIR arms. The X-Shooter spectra were directly downloaded from the ESO Phase 3 archive. The spectra are corrected for barycentric radial velocity based on the observation date.

 We made use of the Gaia ELS catalogue along with their estimated astrophysical parameters to identify new candidates of intense B-type emission stars, as detailed in Appendix \ref{app:A}.
\begin{figure*}
    \centering
    \includegraphics[width=2\columnwidth]{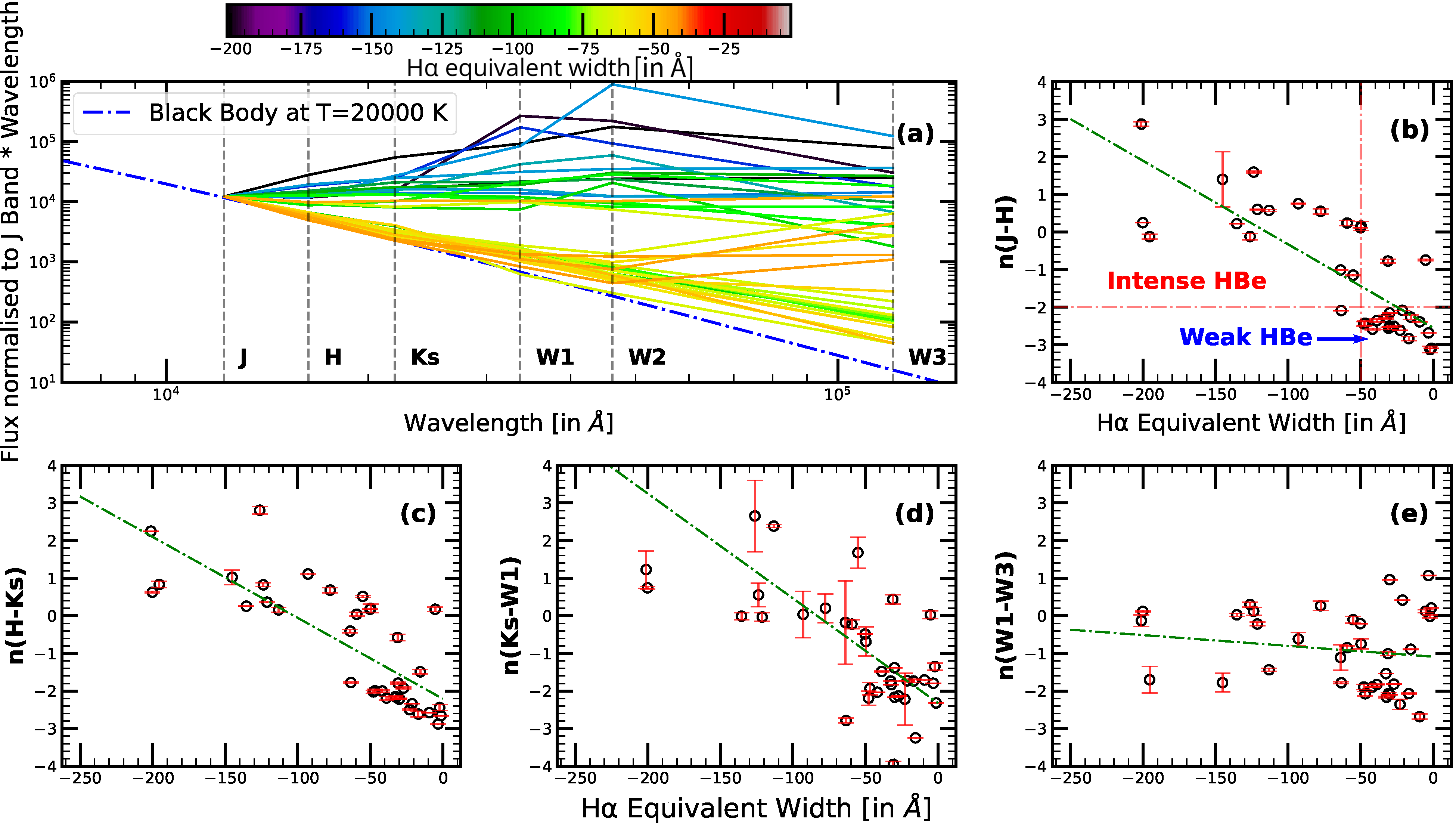}
    \caption{Subplot (a) visualises the NIR excess present in the SED of early HBe stars studied in this work. The colour of each SED represents the H$\alpha$ emission strength. The dash-dotted line represents the blackbody SED for T=20,000 K. The flux values are normalised to J band for visual purpose. The mean wavelength of each IR filter is marked using dashed lines. Subplots (b-e) represent the distribution of Lada indices between IR magnitudes with respect to the H$\alpha$ EW. The green dash-dotted lines in subplots (b-e) show a linear fit to the scatter points shown in each plot. The red dash-dotted lines in Subplot (b) represent the cutoffs (|H$\alpha$ EW| = 50 \AA~and n(J$-$H) = -2) used to separate the intense and weak HBe stars.  }
    \label{fig:1excess}
\end{figure*}

\begin{figure*}
    \centering
    \includegraphics[width=2\columnwidth]{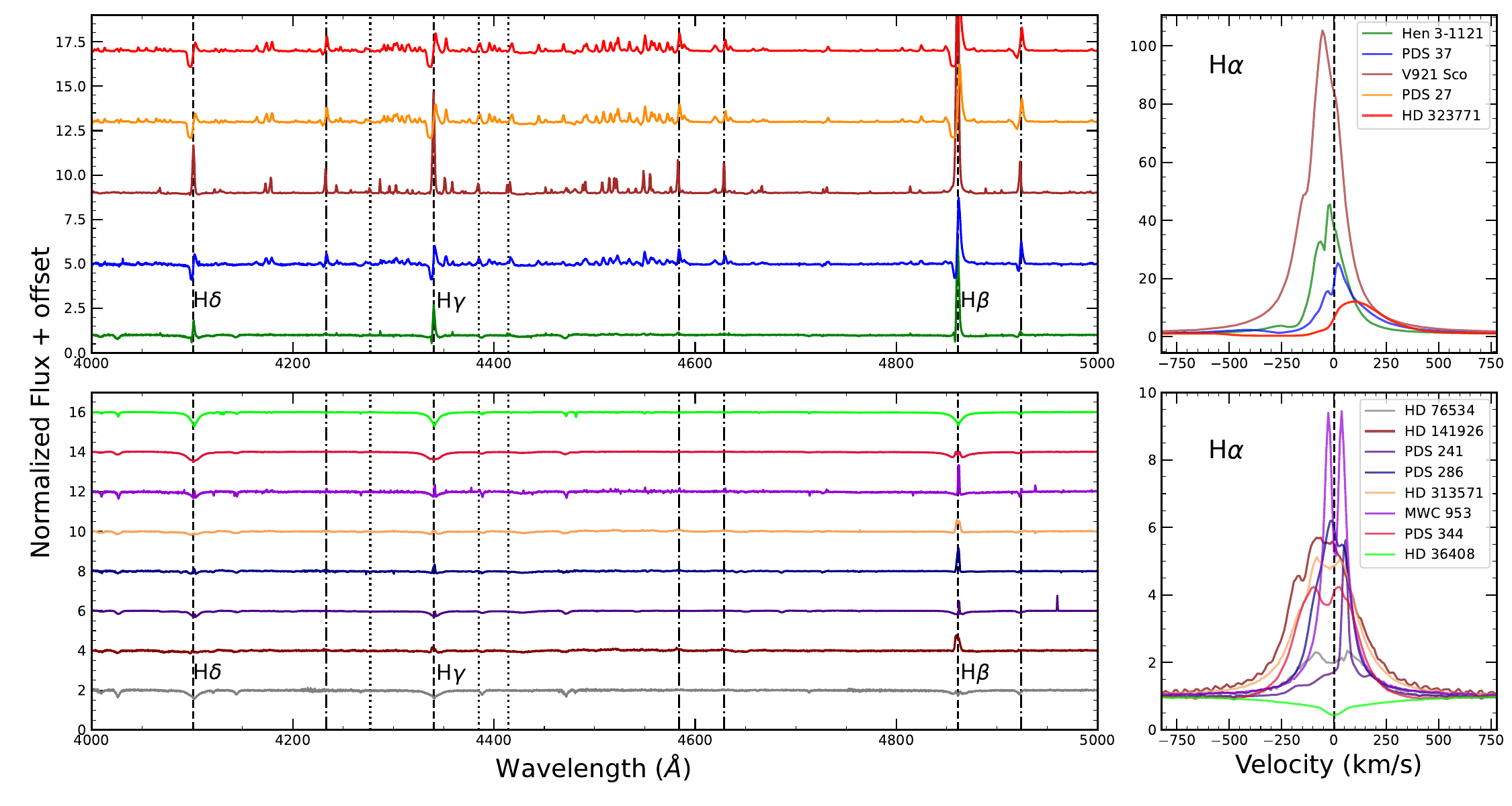}
    \caption{X-Shooter spectra showing the differences in Balmer profiles belonging to intense (top panel) and weak (bottom panel) HBe stars. The top left panel shows the higher order Balmer lines, selected FeII (dash-dotted), and [FeII] (dotted) lines present in the spectra of intense HBe stars. The top right panel shows the H$\alpha$ profiles of intense HBe stars, where the blueshifted absorptions are clearly seen. The bottom panels show the spectra of weak HBes, where the emissions lines are feeble compared to intense HBes. The spectra displayed in this work are not corrected for the doppler shift due to stellar radial velocity.}
    \label{fig:2balmer_xshooter}
\end{figure*}

\section{Analysis and Results}

In this section, we segregate the early HBe stars into two distinct populations based on observed H$\alpha$ EW and NIR excess seen in the SED. Further, we outline the differences in the spectroscopic features between two populations of early HBe stars. 

\subsection{Bi-modality in Infrared Excess and H$\alpha$ EW space}

Previously, \citet{manoj2006evolution} had shown the explicit dependence of NIR excess on the observed H$\alpha$ EW, which suggested the presence of an inner hot disk in HAeBe stars. With the advent of \textit{Gaia} \citep{prusti2016gaia} and IPHAS \citep{barentsen2014second} surveys, homogeneous mass accretion rate studies have been performed in recent years for a well-categorised sample of HAeBe stars \citep{guzmandiaz2021, vioque2020catalogue,arun2019vizier}. As explained, several indirect pieces of evidence suggest that the late HBe and HAe stars accrete through magnetospheric columns, whereas the MA paradigm may not have an active role in early HBe stars. Since early B-type stars do not have large-scale strong magnetic fields, they accrete through the BL mechanism \citep{lynden1974evolution}. This change in accretion mode for early B-type stars significantly changes the evolution of the circumstellar disk. To decode this, we checked for a correlation between NIR excess and H$\alpha$ EW by segregating them based on the spectral types and found that two distinct populations of early HBe stars exist in the NIR excess and H$\alpha$ EW space. The H$\alpha$ EW for our sample of HBe stars is taken from \citet{vioque2018gaia}. We use the Lada\footnote{The equation defining the Lada index is as follows, \begin{equation}
n_{\lambda_1-\lambda_2} = \frac{log(\frac{\lambda_2F_{\lambda_2}}{\lambda_1F_{\lambda_1}})}{log(\frac{\lambda_2}{\lambda_1})}
\label{eqn:3}
\end{equation}, where $\lambda_1$ and $\lambda_2$ represent the wavelength of two bands for which spectral index is calculated. $F_{\lambda_1}$ and $F_{\lambda_2}$ represent the extinction corrected absolute flux measured at $\lambda_1$ and $\lambda_2$.
} indices to quantify the observed excess in each wavelength regime \citep{Lada1987star}. The IR magnitudes are corrected for extinction using respective A$_{\lambda}$ conversion coefficients from \citet{wang2019optical}.

\begin{figure*}
    \centering
    \includegraphics[width=2\columnwidth]{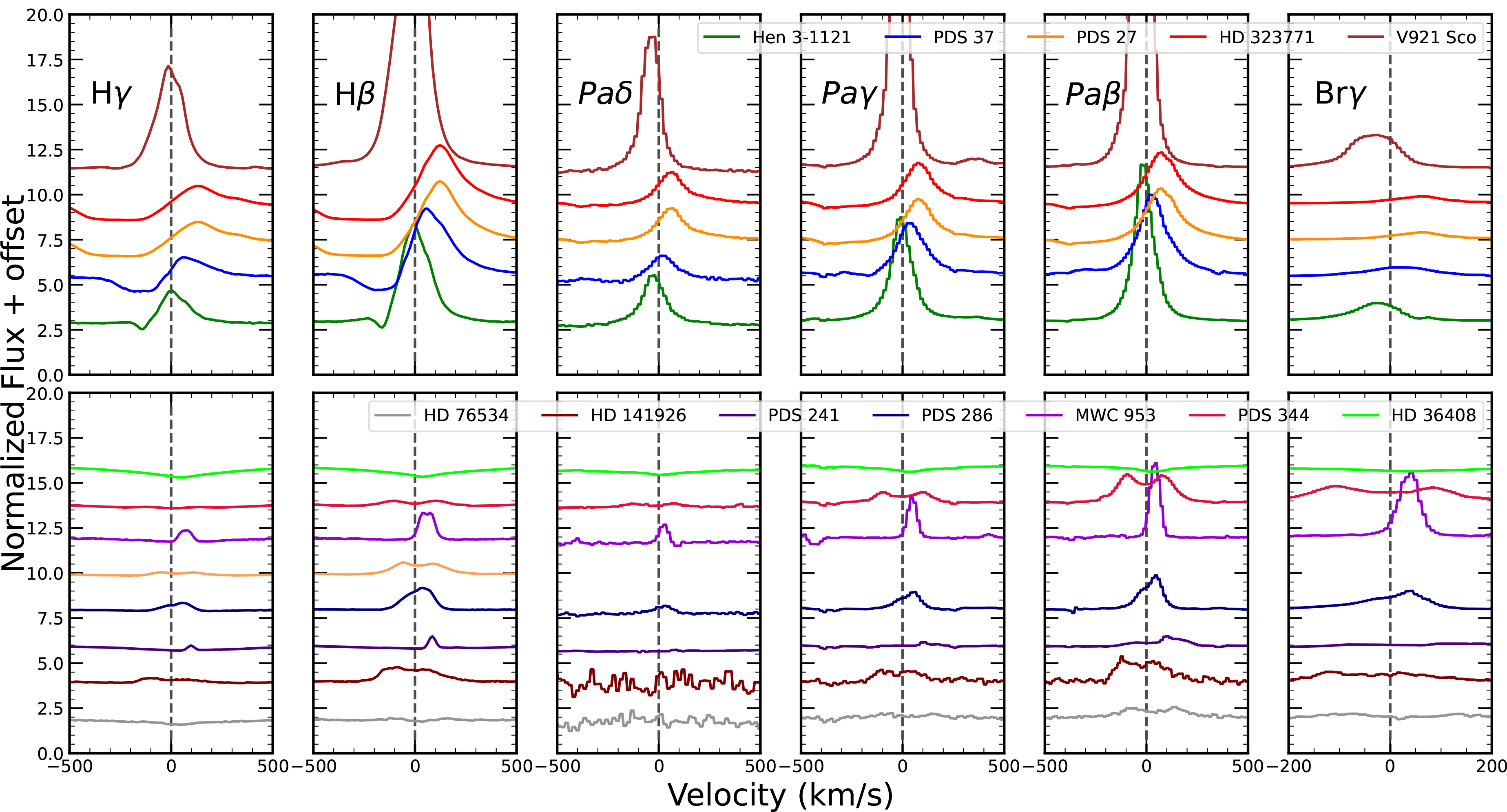}
    \caption{X-Shooter spectra showing the profiles of lower-order H{\sc I} lines belonging to intense (top panel) and weak (bottom panel) HBe stars. The blueshifted absorption feature is seen in the Balmer lines. However, the contribution of blueshifted absorption decreases as we move to Paschen and Brackett lines, denoting the wavelength-dependent opacity of the absorbing medium.}
    \label{fig:2.5balmer}
\end{figure*}

The presence of two sub-populations is visualised in the SED plot (Figure \ref{fig:1excess}a) with the colour of each line denoting the H$\alpha$ EW of the star. For visual purposes, we normalised the flux values of all the stars at the J band. As seen, for stars with high H$\alpha$ EW (> 50 \AA), the SED show excesses at shorter wavelengths (J or H bands). And for stars with low H$\alpha$ EW, the excesses start at longer wavelengths (K$_S$ or more). Even though some outliers show relatively high NIR excess for low H$\alpha$ emission strength, the overall trend of stars with high H$\alpha$ EW having high NIR excess is visible. The presence of two distinct populations is also seen in Figures \ref{fig:1excess}b and \ref{fig:1excess}c, which shows the correlation between NIR indices and H$\alpha$ EW. It is clear from Figures \ref{fig:1excess}a-c, that there exist two different populations of early HBe stars: one with high NIR excess and high H$\alpha$ EW and the other with low NIR excess and low H$\alpha$ EW in the sample of early HBe (B0-B5) stars. Since NIR continuum excess is primarily from the inner hot disk, an intense NIR emission and its correlation with H$\alpha$ emission suggests the presence of an inner hot disk. It is still possible that HBes may have a high MIR/FIR excess caused by the presence of a dusty outer disk. It should be noted that the MIR excess, evaluated using the n(K$_S-$W1) and n(W1$-$W3) indices (Figure \ref{fig:1excess}d,e), does not distinguish the two populations that is seen in NIR--H$\alpha$ EW space. It is pertinent to note that the two populations discussed here only differ in the NIR wavelength region and show similar indices in MIR region, which suggests a similarity in the outer disk of these stars.
For further analysis, we separate the sample of 37 HBe stars into "intense" and "weak" HBe stars based on the cutoffs shown in Figure \ref{fig:1excess}b. The extinction-corrected NIR indices lie close to the photospheric colours for stars with |H$\alpha$ EW| < 50 \AA. We separate the two populations based on median values of the H$\alpha$ EW and n(J$-$H) Lada index distributions. The cutoffs used to distinguish intense and weak HBes are |H$\alpha$ EW| = 50 \AA~and n(J$-$H) = -2. Hence, 15 HBe stars satisfying |H$\alpha$ EW| > 50 \AA~and n(J$-$H) > -2 are classified as "intense HBe" stars and 18 HBe stars with |H$\alpha$ EW| < 50 \AA~and n(J$-$H) < -2 are classified as "weak HBe" stars. The remaining four stars did not fit under these definitions and hence were removed from further analysis. The basic stellar parameters of 15 intense and 18 weak HBe stars are provided in Appendix Table \ref{table2}.

It should be noted that non-simultaneity of observation exists between H$\alpha$ EW, 2MASS and WISE observations. However, this would not affect our analysis as HBe stars are not highly variable compared to HAe stars (refer to Fig. 8 of \citet{vioque2018gaia}). Moreover, the role of A$_V$ is critical in estimating the NIR excess of the Herbig stars. However, as discussed in section 4.1, the uncertainty in the A$_V$ value does not affect the presence of two sub-populations in NIR excess vs H$\alpha$ space.

\begin{figure*}
    \centering
    \includegraphics[width=2\columnwidth]{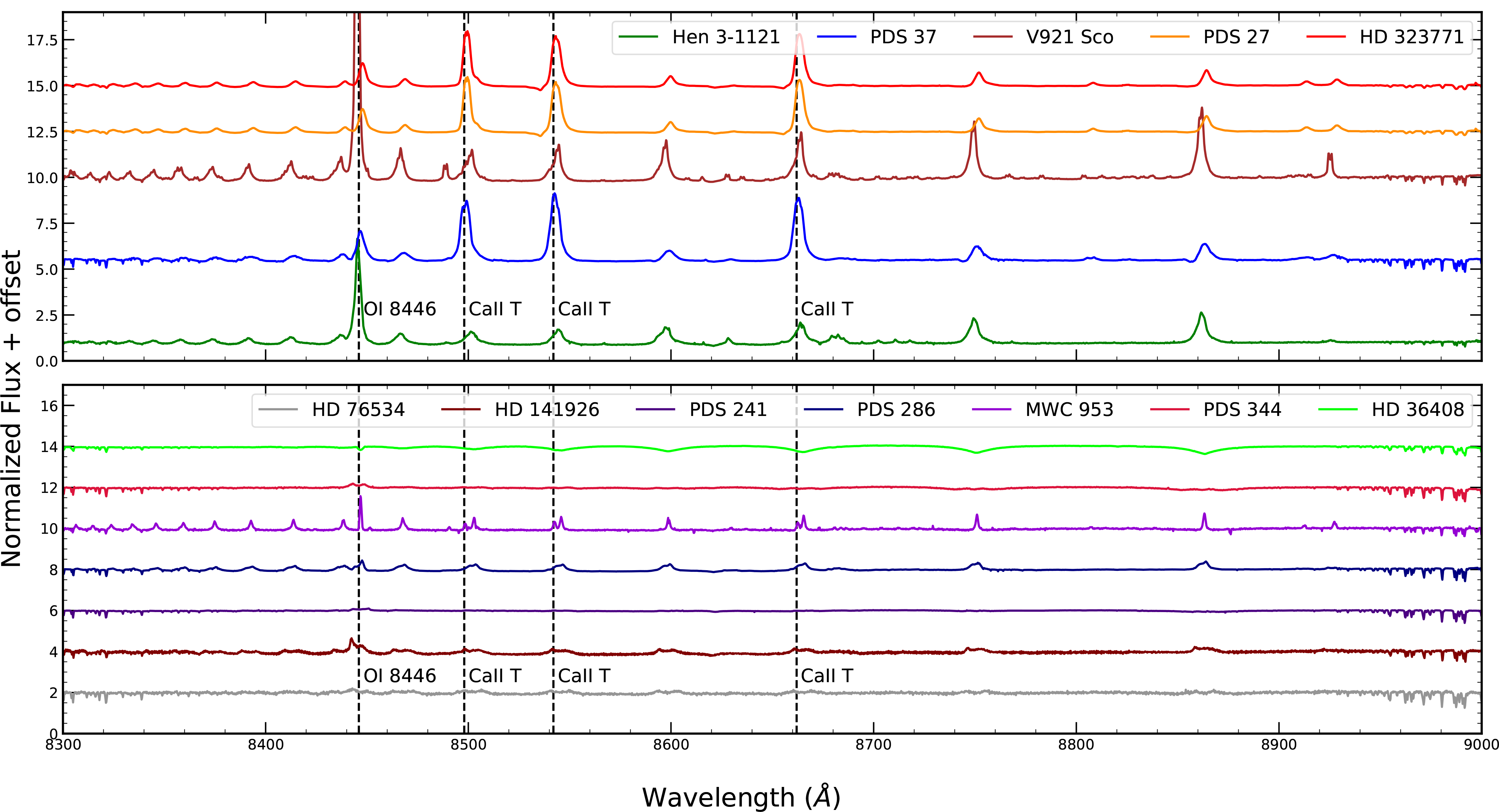}
    \caption{X-Shooter spectra showing the differences in Paschen and Ca{\sc II} triplet profiles belonging to intense (top panel) and weak (bottom panel) HBe stars. The O{\sc I} 8446 \AA~and Ca{\sc II} triplet lines are marked with dashed lines. The difference in intensity of the emission lines can be seen between the top and bottom panels. }
    \label{fig:3paschen}
\end{figure*}

\begin{figure*}
    \centering
    \includegraphics[width=2\columnwidth]{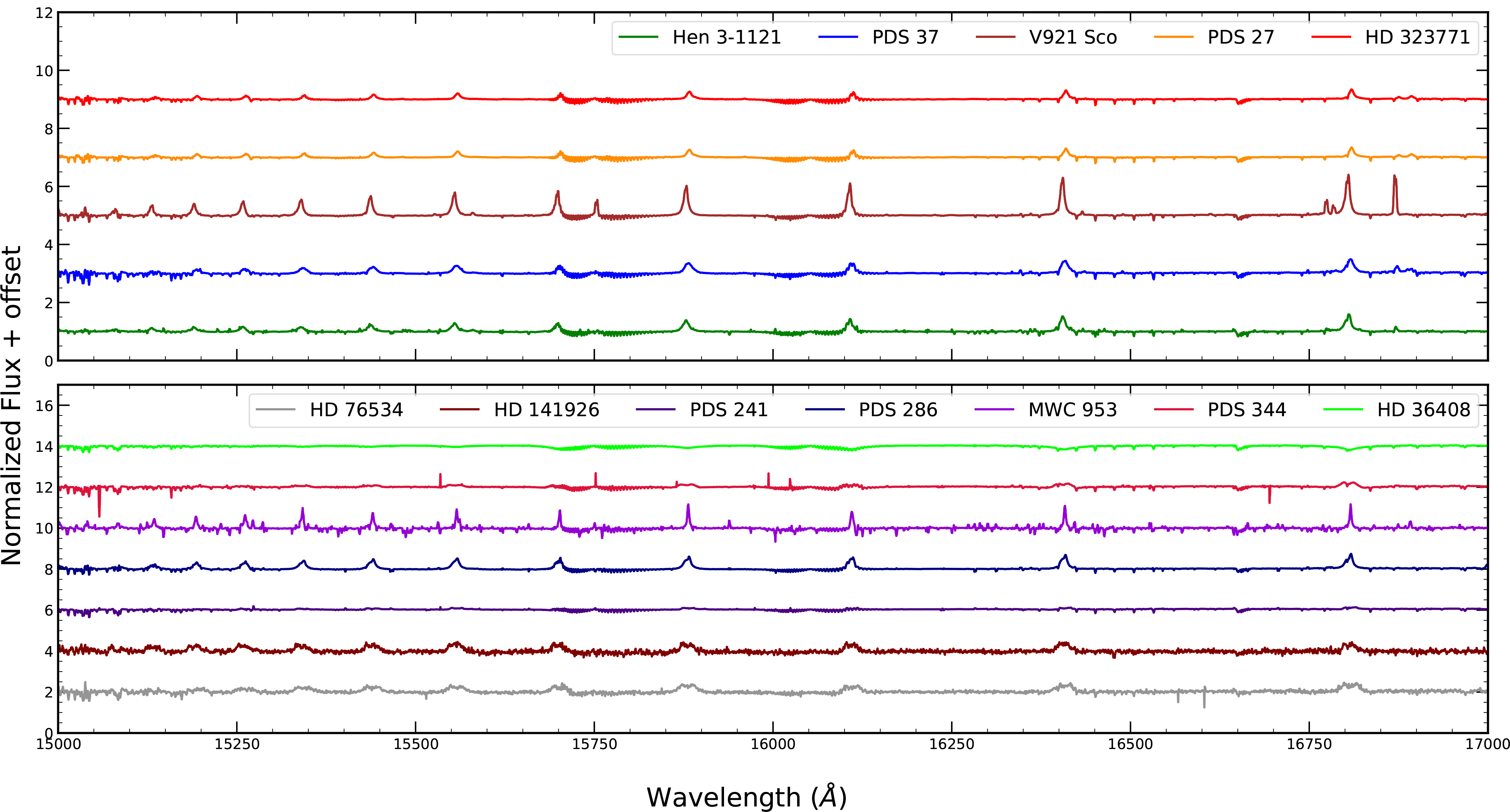}
    \caption{X-Shooter spectra showing the differences in higher-order lines of Brackett series belonging to intense (top panel) and weak (bottom panel) HBe stars. It is interesting to see that the higher-order Brackett series lines are of comparable intensities between intense and weak HBes in contrast to all other emission lines. }
    \label{fig:4brackett}
\end{figure*}

\begin{figure*}
    \centering
    \includegraphics[width=2\columnwidth]{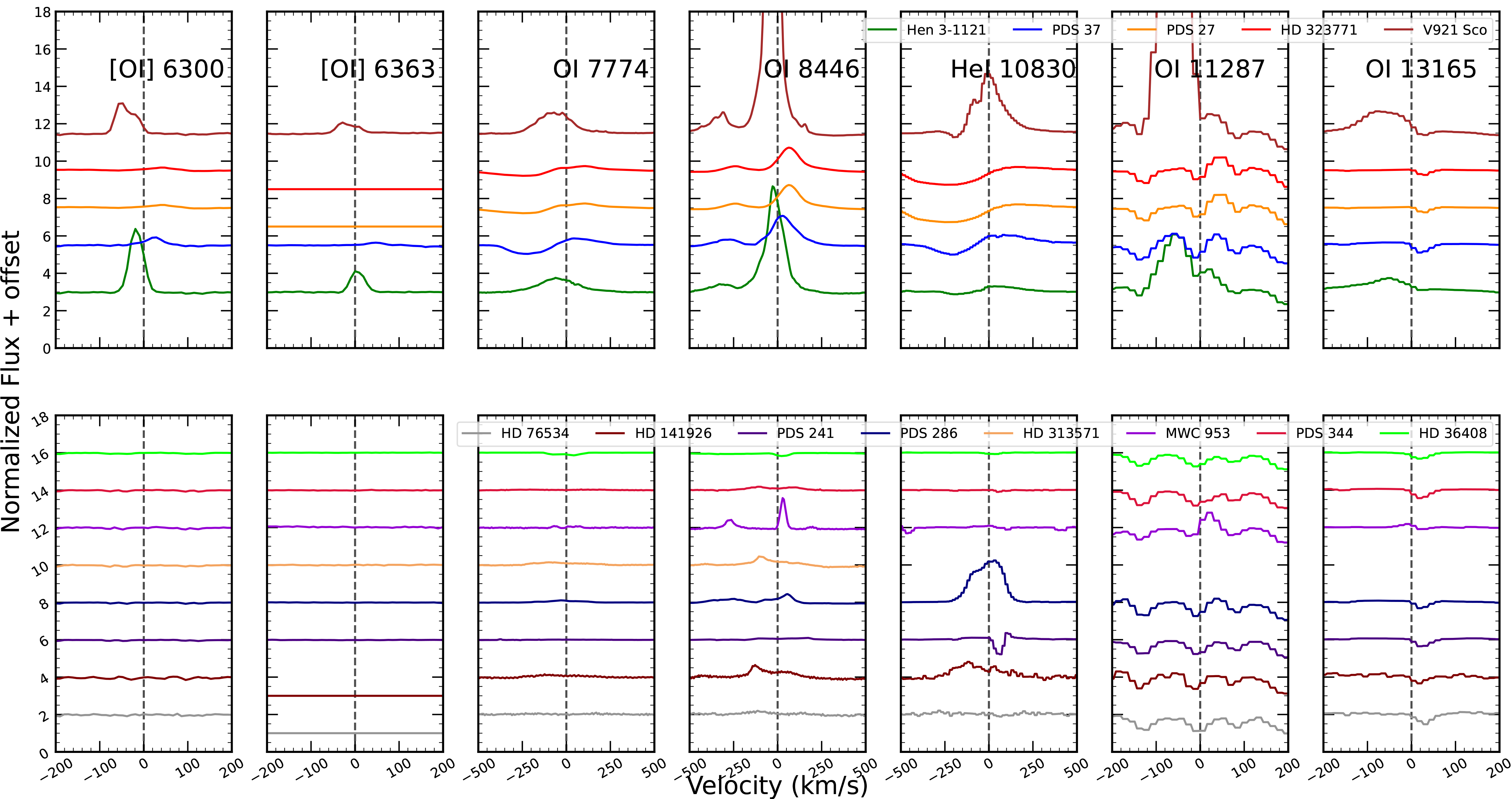}
    \caption{Plot of X-Shooter spectra showing the differences in metallic lines belonging to intense (top panels) and weak (bottom panels) HBe stars. The stark contrast between intense and weak emitters is clear based on the presence of O{\sc I} 7774 \AA, 8446 \AA, 11287 \AA~and 13165 \AA~lines. The He{\sc I} 10830 \AA lines of intense HBe stars show blue-shifted absorption, similar to their Balmer profiles. }
    \label{fig:5oi}
\end{figure*}


\subsection{Difference in the spectral features of HBe stars belonging to two subpopulations}

In addition to the differences in NIR excess and H$\alpha$ EW of the two sub-populations, there exist clear differences in various emission lines that are often found in HAeBe stars. Based on the presence of higher-order H{\sc I} emission lines in Herbig stars, \citet{shridharan2023hi} pointed out that some young stars lose the dynamic inner circumstellar medium very early in their PMS phase. The difference in the circumstellar medium manifests in the presence/absence and the intensity of spectral lines. Hence, we evaluate the distinction in the spectral lines/features using the VLT/X-Shooter spectra. Only 13 out of 37 HBe stars in our list have X-Shooter spectra. To maintain the homogeneity in the analysis, we use only the available X-Shooter spectrum and do not consider spectra from other instruments. We plan to perform a spectroscopic survey of all 37 HBe stars in the future, which is beyond the scope of this work. This section details differences in various spectral features observed in the X-Shooter spectrum belonging to five intense and eight weak HBe stars. 
\subsubsection*{H{\sc I} emission lines}
Figure \ref{fig:2balmer_xshooter} shows the differences in Balmer emission lines of intense and weak HBe emitters. Over the years, it has been clear that the H$\alpha$ profile, especially in the case of YSOs, is a combination of emission through different mechanisms such as accretion, disk, and/or winds. Furthermore, it is susceptible to optical depth effects \citep{kurosawa2006halpha}. We see that (in the top right panel of Figure \ref{fig:2balmer_xshooter}) the H$\alpha$ profiles of intense emitters are not symmetrical and show weak blueshifted absorption in all the cases. This is different from the profiles of weak emitters, in which most of them show symmetrical double peaks and only one (HD 141926) shows blueshifted absorption.  These profile variations also indicate the differences in the inner circumstellar environment of HBe stars, similar to what we assessed from the SED analysis. The blueshifted absorptions are seen at velocities of $\sim$150-300 km/s, similar to the wind velocities reported in \citet{cauley2015optical}. These differences are not only seen in H$\alpha$ but also in H$\beta$ and H$\gamma$ lines (Figure \ref{fig:2.5balmer}). The detection of blueshifted absorption features and non-detection of redshifted absorption features in HBe stars suggests the possibility of BL being the favourable accretion mechanism \citep{cauley2015optical}.

In addition, we explored the advantage of X-Shooter spectra to study the profiles of Paschen and Brackett lines observed during the same epoch, which are shown in Figures \ref{fig:3paschen} and \ref{fig:4brackett}, respectively. We see that the blueshifted features are not prominent in Pa$\beta$, Pa$\gamma$ and Br$\gamma$ lines (Figure \ref{fig:2.5balmer}), suggesting that these features can be wavelength dependent and/or due to the contribution from different emission regions. However, the blueward asymmetry is still present in the lower-order lines of the Paschen and Brackett series. It should be noted that in the case of higher-order lines of the Paschen series, the asymmetry in the profiles can be seen to an extent. However, the asymmetry completely disappears for higher-order lines of the Brackett series. This again points to the wavelength-dependent opacity of the medium, causing the blueshifted absorption. Interestingly, there is no distinction between intense and weak emitters when it comes to the presence of higher-order lines of the Brackett series (Figure \ref{fig:4brackett}). 

\subsubsection*{Fe{\sc II} emission lines}
HAeBe stars show emission lines belonging to metallic species such as Fe{\sc II}, Ca{\sc II} triplet, O{\sc I} and [O{\sc I}] lines in their spectra \citep{hamann1992emission}. We also observe differences in intense and weak emitters based on the presence of these metallic lines. Fe{\sc II} lines (particularly those present in the UVB arm of X-Shooter spectra) are seen in emission in intense HBe stars, whereas they are completely absent in weak emitters (Figure \ref{fig:2balmer_xshooter}). Since Fe{\sc II} lines (especially 4924 \AA) are seen in P-Cygni morphology in three of five intense emitters, they may arise from a region of outflow. Furthermore, many weak emission lines in the wavelength range 4350--4650 \AA~of intense HBe stars are completely absent in weak HBe stars. 

\subsubsection*{Ca{\sc II} emission lines}
Ca{\sc II} triplet lines (8498 \AA, 8542 \AA, and 8662 \AA) require lower energy for ionisation and can form in cooler regions of the circumstellar disk. Ca{\sc II} triplet lines are seen in three (out of five) intense emitters and in one (MWC 953) of the weak emitters (Figure \ref{fig:3paschen}). \citet{hamann1992emission} studied the Ca{\sc II} triplet emission in HAeBe stars and noted that 71\% of the HBe stars show Ca{\sc II} triplet in emission. They propose that the Ca{\sc II} triplet emission should come from denser regions of the hot star. The low-ionisation energy (Energy of the upper level in the transition [E$_{upper}$] $\sim$ 3.1 eV) of the triplet means they arise from cooler regions away from the star and may not be associated with the inner hot gaseous disk. Further, the observed ratios of Ca{\sc II} triplet emission is 1:1:1 instead of $\sim$1:10:6, as expected by the ratio of oscillator strengths (log(gf))\footnote{https://physics.nist.gov/PhysRefData/ASD/lines\_form.html}. Hence they arise from an optically thick region. It is also interesting to note that three stars showing Ca{\sc II} triplet also show Ca{\sc II} doublet emission (8912 \AA~and 8927 \AA). The Ca{\sc II} doublet ratio is also close to unity with 8927 \AA~slightly more intense than 8912 \AA. As discussed in \citet{hamann1992emission2}, the Ca{\sc II} may be arising due to greater saturation of triplet lines caused by the optical thickness of the Ca{\sc II} emitting medium. The difference in energy levels of Ca{\sc II} triplet (E$_{upper}$ $\sim$ 3.1 eV) and doublet (E$_{upper}$ $\sim$ 8.4 eV) lines points to different regions of line formation. Interestingly, a weak HBe star, MWC 953, shows both Ca{\sc II} triplet and Ca{\sc II} doublet lines in emission (although weak compared to intense emitters).

\subsubsection*{O{\sc I} emission lines}
\citet{mathew2018excitation} showed that the most likely excitation mechanism for the formation of O{\sc I} lines is Ly$\beta$ fluorescence. They observed that the emission strengths of 8446 \AA~and 11287 \AA~are more intense than the adjacent O{\sc I} lines at 7774 \AA~and 13165 \AA. We observed O{\sc I} lines in emission in all five intense HBe stars and one (MWC 953) of the weak emitters (Figure \ref{fig:5oi}). It should be noted that Ca{\sc II} triplet emission and O{\sc I} lines seen in MWC 953 are weaker and narrower compared to the lines observed in intense emitters. For intense HBe stars, V921 Sco and Hen 3-1121, the ratio of the observed pair of lines (8446/7774 and 13165/11287) follows the ratio observed in \citet{mathew2018excitation}. The observed [log(F(8446)/F(7774)), log(F(11287)/F(13165)] ratios are [0.79, 0.68] and [0.59, 0.74] for V921 Sco and Hen 3-1121, respectively\footnote{We did not measure ratios for other intense HBe stars since O{\sc I} lines show p-cygni profile, which can affect the calculation}. We also note that the intensity of O{\sc I} directly correlates with the spectral type. The intensity is maximum for V921 Sco (B0-B1), which falls off for Hen 3-1121 (B1-B2) and eventually becomes negligible for PDS 37 (B2-B3). The O{\sc I} lines (especially the IR lines) are non-existent for PDS 27 and HD 323771 (B4-B5). Considering [O{\sc I}] lines, which are used as wind indicators in HAeBe stars \citep{corcoran1998wind}, we see that V921 Sco, PDS 37, and HD 323771 show both 6300 \AA~and 6363 \AA~lines. The rest of the stars studied do not show [O{\sc I}] in emission.

\subsubsection*{He{\sc I} 10830 \AA~emission feature}
More importantly, the meta-stable He{\sc I} 10830 emission line is used as a probe to study the dynamics of mass flows around HAe/Be stars\citet{cauley2014diagnosing}. The blueshifted features observed in H{\sc I} lines are also seen in the He{\sc I} 10830 \AA~line of intense HBe stars (Figure \ref{fig:5oi}. The blueshifted ($\sim$ 150-300 km/s) He{\sc I} 10830 \AA~line mimicking the blueshifted feature in Balmer profiles shows that the absorption mechanisms in both cases can be physically related. Hence, the blueshifted absorption feature in Balmer lines may come from stellar wind in the inner circumstellar region or an expanding envelope along the line of sight to the star, which is missing in weak HBes. 

\section{Discussion}  

\subsection{Role of A$_V$}

\begin{figure}
    \centering
    \includegraphics[width=1\columnwidth]{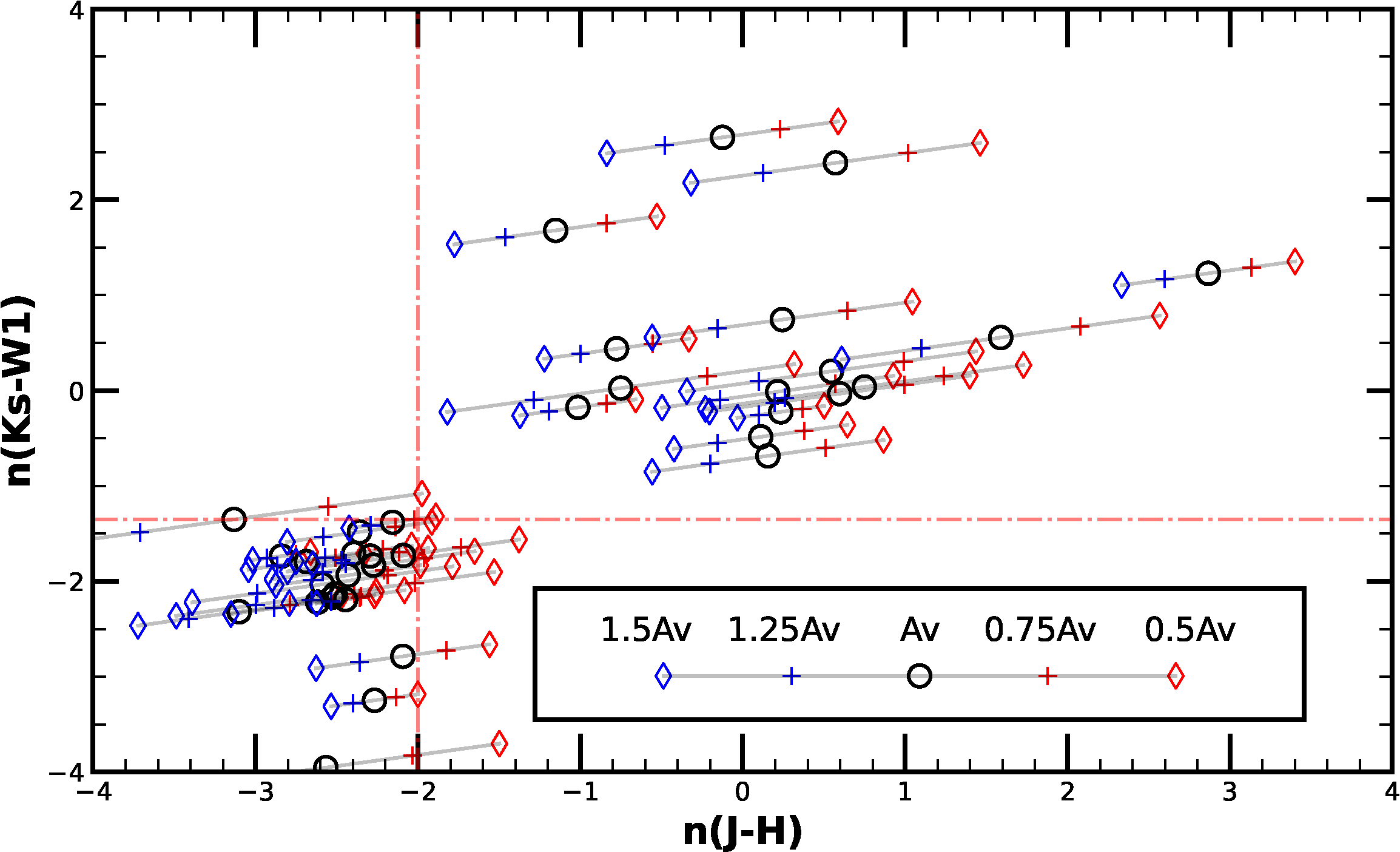}
    \caption{A scatter plot of Lada indices n(J$-$H) and n(K$_S-$W1) for the HBe stars studied in this work. The A$_V$ is varied by 25\% and 50\%, and the resultant index values are marked in 'plus' and 'diamond' symbols, respectively. The symbols are marked in blue if the A$_V$ is increased and in red if the A$_V$ is decreased. The inset subplot explains the symbols and colors used in the plot with the respective A$_V$ value above each symbol. }
    \label{fig:6av_shift}
\end{figure}

Since B-type stars evolve rapidly and reach the main sequence faster than A-type stars, HBe stars are more likely to be associated with the pre-natal cloud. Hence, the line-of-sight extinction can be higher towards the early-type stars. The estimation of A$_V$ towards the early B type depends on identifying accurate spectral type using high-resolution spectra. Hence, the A$_V$ estimation will always carry an inherent uncertainty. To ensure that the NIR excess estimates are not influenced due to the underestimation or overestimation of the A$_V$ value, we varied the A$_V$ by 25\% and 50\% to see the change in the indices. Figure \ref{fig:6av_shift} shows the change in indices if the A$_V$ is varied. We can see that even if the A$_V$ is varied by 50\%, the Lada indices do not vary significantly. The pattern of two populations in n(J$-$H) is still valid even if the A$_V$ is under/overestimated by 50\%. Hence, the A$_V$ value used in this study does not affect the identification of two sub-populations in early HBe stars. The NIR excess seen in intense HBe stars is genuine and is not affected by the uncertainty in the A$_V$ parameter.

\subsection{Bimodality in accretion rates}

There are important studies (e.g. \citealp{mendigutia2011accretion}, \citealp{arun2019vizier} and \citealp{wichittanakom2020accretion}) which have estimated the accretion rates of HBe stars homogeneously using H$\alpha$ EW values. Using the correlation between accretion rates and stellar mass, \citet{wichittanakom2020accretion} showed that the accretion mode changes at around 4 M$_\odot$. It also can be seen from their analysis (Figure 7(right) of \citet{wichittanakom2020accretion}) that there exists a large scatter in the mass accretion rates of Herbig Be stars. Although there exists no clear distinction, we can see that the range of mass accretion rates can vary from 10$^{-7}$ to 10$^{-3}$ M$_{\odot}$ yr$^{-1}$. However, as mentioned, the H$\alpha$ emission line has many contributions and suffers from opacity issues. As seen from the H$\alpha$ profiles of intense HBe stars, there is a blueshifted absorption component, which reduces the total H$\alpha$ strength and underestimates the mass accretion rate. It is possible to overcome the opacity issue using the Br$\gamma$ line, as proposed by \citet{grant2022tracing}. As pointed out by them, there is a change in slope of log ($ {\dot{M}_{acc}} $) vs log($M_\star$) plot for high-mass objects when compared with low-mass HAes. Though there is no direct sample overlap between our work and \citet{grant2022tracing}, we point to Figure 11 of \citet{grant2022tracing}. The figure mimics our finding that two different populations exist in mass accretion rate values for stars belonging to the B0-B5 spectral range. The $ {\dot{M}_{acc}} $ values in \citet{grant2022tracing} for early HBe stars ranges from 10$^{-7}$ to 10$^{-3}$  M$_{\odot}$ yr$^{-1}$. One set of early HBe stars have ${\dot{M}_{acc}}$ in the range of 10$^{-4.5}$ to 10$^{-3}$ M$_{\odot}$ yr$^{-1}$ and another set of stars have values in the range 10$^{-6}$ to 10$^{-7}$ M$_{\odot}$ yr$^{-1}$. Under the assumption that Br$\gamma$ emission directly correlates with the mass accretion rate, the two to three orders of magnitude difference in the mass accretion rates between the two populations is evidence of intense HBe stars undergoing accretion through BL and that the active accretion phase has ended in weak HBe stars. 

\subsection{Spectral features observed in other spectral ranges}

The interesting spectral features seen in intense HBe stars, such as FeII and [FeII] lines in the UVB band of X-Shooter spectra, are not limited to this spectral type. We checked all the X-Shooter spectra of Herbig stars to see the incidence of intense FeII and [FeII] emission lines. We note that stars belonging to B5-B9 (HD 259431, PDS 133 and HD 85567) and A0-A5 (V380 Ori and Z CMa) also show spectral features similar to intense HBe stars. Interestingly, some of these stars have been studied in detail using various techniques. The stars V380 Ori and Z CMa are known to be associated with Herbig-Haro outflows and bipolar jets \citep{reipurth2013hh, whelan20102008}. HD 85567 and HD 259431 stars are shown to have an inner compact gaseous disk of size <1 au \citep{kraus2008detection, bagnoli2010inner, wheelwright2013hd}. Even though they belong to later spectral types, the commonality between the spectral features could mean that the properties of the emission medium can be identical. This can be considered as indirect evidence for the presence of inner gaseous disks and intense outflows in the sample of intense HBe stars. 

\subsection{Comparison with other types of Be stars}

A natural question with the classification of weak HBe stars is whether they truly belong to the Herbig category. Given the mass range of B0-B5 stars (5 - 18 M$_{\odot}$), they typically stay in the PMS phase for less than 2 Myr (MIST; \citealp{dotter2016mesa, choi2016mesa}). Hence, finding them in the PMS phase is difficult due to the rapid evolutionary timescales. Thus, the validity of these massive stars being in the Herbig phase is always under scrutiny.

Since the photometric and spectroscopic signatures of weak HBes and Classical Be (CBe) stars are very similar, CBe stars may be masquerading as the weak HBes in our sample. Similarly, over the years, stars such as MWC 137, classified as young HBe stars, turned out to be evolved supergiant stars by recent studies \citep{kraus2021resolving}. In the context of intense HBe stars, it is essential to mention the stars classified in the literature as "B-type stars showing B[e] phenomenon". B[e] phenomenon was described by \citet{lamers1998improved} as B-type stars showing high NIR excess and intense H$\alpha$ emission along with allowed and forbidden (especially [O{\sc I}], [FeII] and [NII]) emission lines in their spectra. The B[e] phenomenon does not occur at a particular evolutionary stage. Instead, it has been observed in pre-main sequence stars, supergiants, Symbiotic binaries, and Planetary Nebulae (PNe). B[e] stars are highly reddened, and their spectra do not show any photospheric lines. It has been shown by \citet{de1998search} that stars showing B[e] are not part of young clusters or nebulosity. The similarity in spectral and NIR characteristics of our intense HBes and B[e] stars can affect the classification. Hence, the validity of the young nature of our HBe sample needs to be rechecked. In this section, we assess the information available in the literature to confirm the HBe nature of our sample of stars.

\textit{Intense HBe stars}: Hen 3-1121 is confirmed to be a young HAeBe star by \citet{carmona2010new} through high-resolution spectroscopy. PDS 37 and 27 are studied in detail to be massive YSO stars undergoing intense accretion (10$^{-3}$ to 10$^{-4.5}$ M$_{\odot}$ yr$^{-1}$) and are on track to become O-type stars. They are observed to have flattened inner disk structure \citep{ababakr2015spectroscopy}. V921 Sco was first classified as a Bep supergiant star by \citet{hutsemeker1990}. However, due to the non-detection of the CO bandhead and the observed emission region of Br$\gamma$, it is classified as a pre-main sequence star by \citet{kraus2012nature}. There is no relevant literature about HD 323771 star, except that it belongs to B5V spectral type \citep{vieira2003}. 

\textit{Weak HBe stars}: There has been considerable ambiguity regarding the evolutionary phase of HD 76534 in the literature. It belongs to Vela R2 association \citep{herbst1975}, which is a young star formation region, and the star has a nebular region associated with it. Recently, \citet{patel2017photoionization} have modelled the circumstellar disk and have shown the presence of a passive, optically thick disk. From interferometric studies, it has been suggested that HD 141926 possibly accrete through the BL mechanism \citep{marcos2021k}. PDS 241 was suggested to be CBe star due to its low MIR flux \citep{verhoeff2012mid} but has been studied as a HAeBe star in the recent works of \citet{grant2022tracing} and \citet{guzmandiaz2021}. HD 313571 and MWC 953 are confirmed as HAeBe stars by high-resolution optical spectroscopy \citep{carmona2010new}. They are also known to be associated with nebulosity in Spitzer images. There are no relevant literature about the nature of PDS 286 and HD 36408. 

It should be noted that there is no distinction between the intense and weak HBe stars based on age, mass and luminosity values (based on the values listed in \citet{guzmandiaz2021}). Hence, it could be said that the intense and weak HBe stars belong to similar evolutionary phases. However, we are aware of the caveat here that the characteristics of intense HBe stars such as, intense H{\sc I} emission, [Fe{\sc II}] emission and blue-shifted absorption features in H{\sc I} lines are observed in supergiant B[e] (sgB[e]) stars as well \citep{lamers1998improved}. A fail-safe method to distinguish HBe and sgB[e] stars is through the detection of the $^{13}$CO bandhead emission and ratio of $^{12}$CO/$^{13}$CO features \citep{kraus2009pre}. Hence, high-resolution K-band spectroscopy is required to confirm the validity of the young nature of the intense HBe stars. In addition, V921 Sco is identified to have a very close ($\sim$0.025") late B-type companion star through interferometry \citep{kraus2012naturea}. They suggest that the B[e] phenomenon observed in V921 Sco is consequence of binary interactions, where the forbidden-line emitting material gets ejected during the episodic interaction phases. Thus, it could be possible to explain the features of intense HBe stars through interaction with their undetected close binary companions. To explore the possibility of close undetected companions and to confirm the evolutionary phase of intense HBe stars, we plan to perform high angular resolution interferometry and high-resolution K-band spectroscopy in the near future. Since previous studies indicate that most of the intense and weak HBe stars are found to be young and associated with the nebulosity, we treat them as young PMS stars in this work.

\section{Summary and Conclusions}

In this work, we made use of the archival photometric and spectroscopic data available for early HBe stars to study the two sub-populations within the sample of early HBe stars. We analyse the correlation between NIR excess and H$\alpha$ emission strength, supported by the distinction in the spectral features, to show that the inner circumstellar medium belonging to intense and weak HBe stars are different. Though there have been indications for two sub-populations of early HBe stars, this is the first work to point out the clear differences between them using IR photometry and various spectroscopic features using VLT/X-Shooter spectra. The major results are summarised below, 
\begin{itemize}
    \item The NIR excess, as quantified by the n(J$-$H) Lada index, shows that one population of HBe stars have weak NIR excess, whereas the other population of HBe stars have very high NIR excess. Since the NIR excess corresponds to the inner circumstellar medium of the Herbig star, the dichotomy between the populations also corresponds to the difference in the evolution of the inner circumstellar medium. 
    \item Furthermore, the stars with high NIR excess also seem to show higher H$\alpha$ EW and vice-versa. The emission strength of H$\alpha$ has always been directly used to estimate the mass accretion rate in PMS stars. The high H$\alpha$ EW values, thus also correspond to higher rate of accretion in these stars.  
    \item Since the early HBe stars are known to accrete through the BL mechanism, we propose that the difference in H$\alpha$ and NIR excess indirectly points to two types of early HBe stars -- those HBe stars with an inner disk and accreting through BL mechanism, and a second population with an evolved circumstellar medium where the BL accretion has stopped.
    \item This is also supported by the differences in the spectral features seen in X-Shooter spectra for a subsample of HBe stars. Most importantly, the H$\alpha$ profiles of all intense HBe stars show blueshifted absorption features. However, in the weak HBe stars, the H$\alpha$ profiles are symmetrical. 
    \item The blueshifted absorption in intense HBe stars is also seen in other HI lines such as H$\beta$, Pa$\beta$, and Br$\gamma$. It should be noted that the blueshifted absorption is not relatively well pronounced in Brackett and Paschen lines, but the blueward asymmetry of the profiles is clear. 
    \item The metallic lines, O{\sc I}, [O{\sc I}], and CaII triplet often seen in HAeBe stars also show the distinction between intense and weak HBes. O{\sc I} 7774 \AA, 8446 \AA~and [O{\sc I}] 6300 \AA, 6363 \AA~lines are only present in intense HBe and completely absent in weak HBe. Since [O{\sc I}] lines form due to outflows from the star, it can be concluded that there are no stellar winds or outflows from weak HBe stars.
    \item HeI 10830 \AA, a meta-stable transition line, has been studied extensively by \citet{cauley2014diagnosing,cauley2015optical} to estimate accretion and mass flow around Herbig stars. The non-detection of blueshifted HeI 10830 \AA~in weak HBe shows no mass flow from the star. It acts as an indirect indicator of dampened activity in the circumstellar medium of weak HBe stars. The blueshifted velocities of the HeI 10830 \AA~absorption feature also correspond with the blueshifted velocities seen in H$\alpha$ and H$\beta$ profiles in intense HBe stars. 
    \item Many lines belonging to different multiplets of FeII and [FeII] transitions are seen in the wavelength range 4200--4450 \AA~in intense HBe stars and are absent in the weak HBe stars. Another interesting spectral feature we noted in the presence of CaII doublet lines at 8912 \AA~and 8927 \AA~ in three intense HBe stars. 
    \item Most of the stars used in this study are classified as Herbig Be stars from extensive, independent studies. Hence, the possibility of B[e] stars and CBe stars being classified as intense HBe and weak HBe, respectively, can be neglected.  
    \item We have identified 44 intense HBe candidates from the latest \textit{Gaia} DR3 database. A detailed spectroscopic analysis is necessary to confirm the intense HBe nature of these 44 candidates (Appendix \ref{app:A}).
\end{itemize}

\begin{acknowledgements}

We want to thank the Science \& Engineering Research Board (SERB), a statutory body of the Department of Science \& Technology (DST), Government of India, for funding our research under grant number CRG/2019/005380. AS and RA acknowledge the financial support from SERB POWER fellowship grant SPF/2020/000009. The authors are grateful to the Centre for Research, CHRIST (Deemed to be University), Bangalore, for the research grant extended to carry out the current project (MRPDSC-1932). We thank the SIMBAD database and the online VizieR library service for helping us with the literature survey and obtaining relevant data. This work has made use of the ESO Phase 3 science archive facility. 
\end{acknowledgements}

\bibliographystyle{aa}
\bibliography{example}

\begin{appendix}

\section{Identification of new intense HBe candidates from \textit{Gaia} DR3}
\label{app:A}

As shown in previous sections, two distinct populations of massive Herbig stars seem to have drastically different inner circumstellar environments. The works of HAeBe stars over the years have been skewed towards late HBe and HAe stars. Only in the recent works (e.g., \citealp{guzmandiaz2021,mendigutia2011accretion}), effort has been taken to include more HBe stars into the analysis. Using the latest \textit{Gaia} DR3, it is possible to identify new intense HBe candidates. \textit{Gaia} DR3 provides spectroscopic parameters for 2.3 million hot stars and a catalogue of 57,511 emission-line stars based on the pseudo-equivalent width (pEW) measurements. The catalogue provides various ELS classes such as `beStar', `TTauri', `HerbigStar' and `PN' based on their BP/RP spectra. We use the ELS catalogue given by \textit{Gaia} DR3 and 2MASS colours to search for more intense HBe candidates.
In order to search for new intense HBe stars from \textit{Gaia} DR3, we queried the ELS catalogue using the \textit{Gaia} ADQL facility. We cross-matched the stars classified as `HerbigStar' in \textit{Gaia} DR3 with the 2MASS catalogue and got 3825 sources. \citet{shridharan2022emission} proposed a second-order polynomial to convert pseudo-EW to observed H$\alpha$ EW of emission-line stars. We used the highest piece-wise fit parameter values in Figure 3 of \citet{shridharan2022emission} to estimate the observed H$\alpha$ EW of the stars. Then, 325 stars with the estimated |H$\alpha$ EW| > 50 \AA~ are retained. To find the hot stars from this sample, we selected those stars with \textit{teff\_gspphot} > 15000 K. The estimate \textit{teff\_esphs} was not used because it was not available for a significant fraction of our candidate objects. After the \textit{teff\_gspphot} cutoff, 65 stars are retained. Further, n(J$-$H) was estimated for these 65 sources using the extinction value provided by \textit{Gaia} DR3. Finally, we have identified 44 potential intense HBe candidates from this analysis. Of 44 candidates, 36 are already recorded in the SIMBAD database. A table containing information regarding the possible intense HBe candidates is provided in Table \ref{tab:a1}. Even though \textit{Gaia} DR3 has classified these stars as `HerbigStar', one should be cautious about accepting this classification. Some stars given in Table \ref{tab:a1} may belong to PNe and supergiant phases. Through this section, we intend to demonstrate the use of large-scale surveys such as \textit{Gaia} to identify interesting set of objects such as intense HBe stars. A detailed and careful spectroscopic study is needed to prove the young nature of these stars. 

\begin{landscape}
\begin{table}
\tiny
\caption{Table containing information about the intense HBe candidates identified from \textit{Gaia} DR3}
\label{tab:a1}
\begin{tabular}{cccccccccccc}
Gaia\_Source\_ID    & RA    & Dec  & Gaia\_ClassELS & A$_G$ & Distance (in pc) & $T_{eff}$(K)  & pEWHa(nm) & EWHa\_conv(\AA) & n(J-H)  & Simbad\_ID       \\ \hline
2065600341322602112 & 313.5307746 & 41.58276107 & HerbigStar & 4.5 & $771_{-9}^{+9}$ & 19,437 & -3.06 & -76.27 & 1.36 & V* V1219 Cyg \\
5864938685013848320 & 202.1618059 & -63.82954127 & HerbigStar & 3.2 & $5076_{-335}^{+162}$ & 15,012 & -13.78 & -331.44 & 1.11 & PN Th  2-B \\
5258349561694308096 & 151.1261185 & -58.66444562 & HerbigStar & 2.9 & $984_{-39}^{+30}$ & 19,161 & -5.34 & -130.47 & 0.98 & HD  87643 \\
2020112789377651712 & 296.0219088 & 23.4466286 & HerbigStar & 3.6 & $2716_{-57}^{+100}$ & 19,503 & -10.7 & -257.98 & 0.7 & Hen 2-446 \\
2170966574983202560 & 322.0225077 & 49.68336095 & HerbigStar & 2.5 & $2345_{-41}^{+176}$ & 15,009 & -2.33 & -58.81 & 0.67 & EM* GGR   25 \\
4661416379669398144 & 77.40419259 & -67.91911637 & HerbigStar & 0.8 & $16807_{-273}^{+289}$ & 17,370 & -6.48 & -157.68 & 0.47 & 2MASS J05093702-6755086 \\
6053784647417889280 & 185.5965192 & -63.28801792 & HerbigStar & 3.0 & $3381_{-946}^{+399}$ & 15,539 & -5.84 & -142.52 & 0.45 & WRAY 16-110 \\
4103969761134412544 & 277.4986 & -14.94720448 & HerbigStar & 2.5 & $2819_{-67}^{+128}$ & 20,803 & -12.43 & -299.19 & 0.31 & SS 389 \\
5868482307854016000 & 200.0148814 & -62.39835807 & HerbigStar & 1.9 & $5836_{-47}^{+48}$ & 15,948 & -2.98 & -74.51 & 0.03 & THA 17-35 \\
5980805834434317184 & 239.5400967 & -53.85511504 & HerbigStar & 3.6 & $4028_{-682}^{+544}$ & 30,636 & -4.72 & -115.69 & -0.01 & WRAY 15-1390 \\
5601822631829205376 & 117.1625314 & -26.67784839 & HerbigStar & 2.2 & $7641_{-4}^{+2}$ & 17,363 & -2.33 & -58.97 & -0.07 & SS 162 \\
4650950785570772864 & 82.87780386 & -71.74670027 & HerbigStar & 1.1 & $12426_{-330}^{+311}$ & 17,340 & -5.59 & -136.55 & -0.07 & SSTISAGEMC J053130.65-714448.2 \\
5972815515955517696 & 260.2314013 & -38.00011208 & HerbigStar & 4.8 & $2078_{-98}^{+94}$ & 28,514 & -6.64 & -161.56 & -0.15 & EM* AS  225 \\
254003305033558784 & 69.64289254 & 46.07931533 & HerbigStar & 4.1 & $2426_{-171}^{+179}$ & 15,155 & -3.51 & -86.9 & -0.18 & 2MASS J04383428+4604454 \\
5308850371274904192 & 144.6439615 & -54.33735311 & HerbigStar & 4.9 & $4127_{-168}^{+43}$ & 19,030 & -5.53 & -135.05 & -0.22 & WRAY 15-413 \\
4256260466614300928 & 277.3570387 & -6.077050094 & HerbigStar & 3.4 & $2356_{-2}^{+2}$ & 23,218 & -4.82 & -118.08 & -0.24 & EM* MWC  300 \\
5972325576126499200 & 258.5750060 & -38.98323981 & HerbigStar & 4.9 & $1606_{-58}^{+154}$ & 36,489 & -5.34 & -130.45 & -0.29 & EM* AS  222 \\
524428838421397760 & 12.52486253 & 64.75973916 & HerbigStar & 3.3 & $2013_{-54}^{+125}$ & 17,637 & -2.87 & -71.7 & -0.3 & EM* GGA    2 \\
5966221298030125056 & 254.7781823 & -42.70234756 & HerbigStar & 4.7 & $956_{-19}^{+53}$ & 36,985 & -7.46 & -180.88 & -0.31 & CD-42 11721 \\
2200793847233509760 & 340.6742594 & 60.40017141 & HerbigStar & 4.4 & $2621_{-22}^{+17}$ & 35,427 & -5.97 & -145.49 & -0.36 & EM* MWC  657 \\
447808893901762176 & 50.16453913 & 56.39948587 & HerbigStar & 5.4 & $2354_{-77}^{+87}$ & 20,563 & -5.58 & -136.33 & -0.37 & IRAS 03168+5613 \\
2207583404553554944 & 349.3996767 & 63.75176856 & HerbigStar & 6.0 & $4080_{-642}^{+342}$ & 36,022 & -6.08 & -148.11 & -0.38 & IRAS 23154+6328 \\
2200017424528999936 & 335.8737099 & 57.63028127 & HerbigStar & 3.9 & $6590_{-294}^{+767}$ & 28,414 & -4.36 & -107.15 & -0.54 & IRAS 22216+5722 \\
4651696391895922688 & 81.94851316 & -71.81464394 & HerbigStar & 1.5 & $18259_{-234}^{+1067}$ & 36,649 & -6.25 & -152.25 & -0.69 & LHA 120-S 165 \\
2017612358184192768 & 359.8332589 & 66.38672132 & HerbigStar & 5.2 & $732_{-9}^{+9}$ & 15,027 & -2.88 & -72.05 & -0.77 & {[}B62{]}  4 \\
4663539300349528192 & 78.33864859 & -65.65405506 & HerbigStar & 1.2 & $18123_{-653}^{+784}$ & 24,524 & -4.49 & -110.33 & -0.77 & SSTISAGEMC J051321.26-653914.6 \\
4066048532745842688 & 271.4125813 & -24.51107999 & HerbigStar & 4.1 & $1365_{-36}^{+48}$ & 15,047 & -4.02 & -99.09 & -0.91 & EM* LkHA  117 \\
4688847309225362944 & 10.27822154 & -73.35296519 & HerbigStar & 0.6 & $19798_{-143}^{+238}$ & 19,658 & -2.25 & -57.08 & -0.93 & OGLE SMC-SC2  61751 \\
5338335905897507456 & 163.5428339 & -60.08733453 & HerbigStar & 2.5 & $8086_{-206}^{+222}$ & 15,007 & -5.82 & -142.0 & -0.99 & 2MASS J10541029-6005145 \\
4661559629674731776 & 74.40327132 & -67.79371587 & HerbigStar & 1.5 & $10414_{-393}^{+410}$ & 32,534 & -7.78 & -188.65 & -1.01 & LHA 120-S  12 \\
5615056663019267584 & 114.7756575 & -24.75136361 & HerbigStar & 2.5 & $3116_{-52}^{+139}$ & 40,742 & -6.03 & -146.95 & -1.13 & CD-24  5721 \\
2015202228698942080 & 347.8139981 & 62.45698608 & HerbigStar & 4.7 & $2277_{-78}^{+108}$ & 15,013 & -2.7 & -67.71 & -1.13 & HBHA 6207-19 \\
4651968177420876160 & 78.0377423& -71.11381847 & HerbigStar & 1.2 & $15904_{-312}^{+497}$ & 16,345 & -2.42 & -61.16 & -1.24 & LHA 120-S 160 \\
3344973478481675264 & 94.68967401 & 15.28117676 & HerbigStar & 4.3 & $2317_{-47}^{+31}$ & 35,001 & -10.69 & -257.9 & -1.38 & EM* MWC  137 \\
4654808696979503744 & 71.48072022 & -70.84493333 & HerbigStar & 2.7 & $15829_{-39}^{+40}$ & 30,005 & -7.88 & -190.97 & -1.45 & UCAC2   1445227 \\
4655158131209278464 & 74.19618924 & -69.84021409 & HerbigStar & 1.3 & $10738_{-130}^{+149}$ & 20,610 & -2.53 & -63.77 & -1.66 & HD 268835 \\
2006219223007452032 & 334.5353473 & 56.09809927 & HerbigStar & 3.5 & $6064_{-525}^{+633}$ & 28,955 & -4.25 & -104.53 & -0.12 & Not listed in SIMBAD \\
5869126656055392000 & 202.9565261 & -60.39146338 & HerbigStar & 4.6 & $5041_{-198}^{+657}$ & 15,019 & -3.13 & -78.0 & -0.25 & Not listed in SIMBAD \\
5863990940342000000 & 201.7669711 & -65.3231748 & HerbigStar & 3.0 & $6916_{-914}^{+557}$ & 15,236 & -3.19 & -79.38 & -0.26 & Not listed in SIMBAD \\
5308151081881494144 & 147.5873831 & -55.01594503 & HerbigStar & 4.7 & $4934_{-936}^{+992}$ & 16,104 & -2.8 & -70.06 & -0.49 & Not listed in SIMBAD \\
5978823075745899520 & 258.3254263 & -34.52832928 & HerbigStar & 3.7 & $8391_{-506}^{+371}$ & 15,005 & -3.23 & -80.25 & -1.37 & Not listed in SIMBAD \\
4090887629972125056 & 273.9313886 & -21.63773383 & HerbigStar & 3.6 & $5762_{-59}^{+82}$ & 31,272 & -6.65 & -161.73 & -1.4 & Not listed in SIMBAD \\
5308789863772681728 & 145.2286581 & -54.59275441 & HerbigStar & 5.2 & $3065_{-544}^{+249}$ & 15,021 & -2.83 & -70.87 & -1.55 & Not listed in SIMBAD \\
2166911954062686976 & 313.7419439 & 47.76434051 & HerbigStar & 3.8 & $5482_{-1715}^{+390}$ & 15,587 & -2.43 & -61.34 & -1.72 & Not listed in SIMBAD

\end{tabular}
\end{table}
\end{landscape}

\section{Table containing the details of early HBe stars studied in this work}

\renewcommand{\arraystretch}{1.3}
\begin{table*}[]
\caption{Table of early HBe stars studied in this work along with the basic parameters such as spectral type, mass and A$_V$ from \citet{guzmandiaz2021}, EW of H$\alpha$ from \citet{vioque2018gaia} and the assigned type (intense/weak) of HBe from this work. n(J$-$H) lada index is calculated using 2MASS photmetric magnitudes.}
\centering
\begin{tabular}{ccccccc}
\hline
Object Name  & Spectral & Mass & Av  & EWHa     & n(J-H)       & Class        \\
&Type&($M_{\odot}$)&(mag)&($\AA$)&&\\
\hline
\hline
CPM 25        & B3-B4    & $6.78_{-1.18}^{+1.05}$    & 4.5 $\pm$ 0.1   & -200.20 & 0.243  & Intense\_HBe \\
HD 200775     & B4       & $7.01_{-0.01}^{+0.09 }$    & 2.0 $\pm$ 0     & -63.83   & -1.015 & Intense\_HBe \\
HD 323771     & B5-B6    & $4.22_{-0.62}^{+0.16}$    & 1.5 $\pm$ 0.19  & -59.35  & 0.234  & Intense\_HBe \\
HD 87643      & B4       & $16.02_{-5.23}^{+3.98}$   & 2.5 $\pm$ 0.17  & -145.20 & 1.397  & Intense\_HBe \\
Hen 3-1121    & B1-B2    & $10.99_{-1.95}^{+1.09}$  & 4.0 $\pm$ 0     & -135.41   & 0.215  & Intense\_HBe \\
Hen 3-938     & O        & $43.92_{-3.99}^{+2.33}$   & 5.5 $\pm$ 0     & -92.95  & 0.749  & Intense\_HBe \\
MWC 1080      & B0       & $17.97_{-2.03}^{+1.91}$    & 5.0 $\pm$ 0.19  & -113.19  & 0.571  & Intense\_HBe \\
MWC 878       & B0       & $18.05_{-0.34}^{+1.17}$     & 3.5 $\pm$ 0     & -55.19   & -1.151 & Intense\_HBe \\
PDS 27        & B4-B5    & $11.65_{-1.6}^{+1.09}$     & 5.0 $\pm$ 0.07  & -77.6    & 0.545  & Intense\_HBe \\
PDS 34        & B4       & $4.31_{-0}^{+0.29}$      & 3.0 $\pm$ 0     & -50.20  & 0.110  & Intense\_HBe \\
PDS 37        & B2-B3    & $9.04_{-0.21}^{+0.73}$      & 5.5 $\pm$ 0     & -123.76  & 1.588  & Intense\_HBe \\
PDS 477       & B2-B3    & $12.17_{-3.74}^{+3.18}$     & 4.5 $\pm$ 0     & -121.19  & 0.596  & Intense\_HBe \\
PDS 581       & B2-B3    & $7.85_{-0.85}^{+0.39}$     & 3.0 $\pm$ 0     & -201.19  & 2.86  & Intense\_HBe \\
V431 Sct      & B2       & $11.21_{-0.21}^{+0.68}$    & 4.0 $\pm$ 0     & -126.20 & -0.124 & Intense\_HBe \\
V921 Sco      & B0-B1    & $23.47_{-3.45}^{+1.11}$    & 5.5 $\pm$ 0     & -195.49  & -0.128 & Intense\_HBe \\
\hline
HBC 7         & B2-B3    & $8.79_{-0.57}^{+0.21}$    & 4.5 $\pm$ 0     & -41.81  & -2.588 & Weak\_HBe    \\
HBC 705       & B2       & $9.28_{-0.3}^{+0.69}$     & 5.5 $\pm$ 0     & -27.11  & -2.509 & Weak\_HBe    \\
HD 141926     & B0       & $16.91_{-0.88}^{+1}$      & 2.5 $\pm$ 0.09  & -46.88   & -2.429 & Weak\_HBe    \\
HD 305298     & O        & $23.22_{-2.94}^{+1.03}$    & 2.0 $\pm$ 0.18  & -3.24    & -2.687 & Weak\_HBe    \\
HD 313571     & B4       & $9.54_{-0.54}^{+0.46}$     & 2.5 $\pm$ 0     & -38.83   & -2.359 & Weak\_HBe    \\
HD 50083      & B3-B4    & $12_{-0.56}^{+0.43}$      & 1.0 $\pm$ 0     & -47.83   & -2.445 & Weak\_HBe    \\
HD 76534      & B2       & $8.61_{-0.54}^{+0.37}$      & 1.0 $\pm$ 0     & -16.84   & -2.839 & Weak\_HBe    \\
Hen 3-1121S   & O-B0     & $19.14_{-3.44}^{+3.92}$    & 3.5 $\pm$ 0     & -1.29    & -3.100 & Weak\_HBe    \\
Hen 3-823     & B4       & $6.28_{-0.65}^{+0.68}$       & 1.5 $\pm$ 0     & -29.83   & -2.525 & Weak\_HBe    \\
IL Cep        & B2       & $10.02_{-0.02}^{+0.1}$     & 3.0 $\pm$ 0     & -22.83   & -2.617 & Weak\_HBe    \\
MWC 655       & B2-B3    & $11_{-0.37}^{+0.52}$        & 1.5 $\pm$ 0.07  & -15.49   & -2.267 & Weak\_HBe    \\
MWC 953       & B4       & $12.06_{-1.22}^{+1.88}$    & 3.5 $\pm$ 0.14  & -32.21  & -2.273 & Weak\_HBe    \\
PDS 286       & B0       & $26.08_{-1.55}^{+1.41}$     & 6.0 $\pm$ 0     & -30.77   & -2.566 & Weak\_HBe    \\
PDS 344       & B5-B6    & $3.56_{-0}^{+0.24}$      & 1.5 $\pm$ 0     & -30.03   & -2.156 & Weak\_HBe    \\
PDS 361S      & B3-B4    & $5.85_{-0.07}^{+0.16}$     & 2.0 $\pm$ 0     & -9.32    & -2.394 & Weak\_HBe    \\
PDS 543       & O-B0     & $26.92_{-2.44}^{+1.07}$    & 6.5 $\pm$ 0     & -2.19    & -3.13 & Weak\_HBe    \\
V361 Cep      & B4       & $5.54_{-0.11}^{+0.06}$      & 2.0 $\pm$ 0     & -32.59  & -2.293 & Weak\_HBe    \\
WRAY 15-1435  & B2-B3    & $9.69_{-0.63}^{+0.31}$      & 4.0 $\pm$ 0     & -21.19   & -2.089 & Weak\_HBe \\
\hline
\end{tabular}
\label{table2}
\end{table*}

\end{appendix}

\end{document}